\documentclass[10pt,aps,prc,twocolumn,superscriptaddress,floatfix,nofootinbib]{revtex4-1}
\usepackage{amsmath,amssymb,mathrsfs}
\usepackage{bm}
\usepackage{ascmac}
\usepackage{geometry}
\usepackage[dvipdfmx]{graphicx}
\usepackage[dvipdfmx]{color}
\usepackage[dvipdfmx,bookmarks=true,bookmarksnumbered=true,bookmarksopen=true,%
                      colorlinks=true,citecolor=blue,linkcolor=blue,urlcolor=blue,%
                      linktocpage=false,setpagesize=false,%
                      ]{hyperref}

\def\0\\{\nonumber\\}

\def\zI{\mathrm{i}\hspace{0.2mm}}
\def\GRAZING{\mbox{\small GRAZING }}

\makeatletter
\newcommand\footnoteref[1]{\protected@xdef\@thefnmark{\ref{#1}}\@footnotemark}
\makeatother

\begin{document}

\title{
Time-dependent Hartree-Fock calculations for multinucleon transfer\\
and quasifission processes in the $^{64}$Ni+$^{238}$U reaction
}

\author{Kazuyuki Sekizawa}
\email[]{sekizawa@if.pw.edu.pl}
\affiliation{Faculty of Physics, Warsaw University of Technology, ulica Koszykowa 75, 00-662 Warsaw, Poland}
\affiliation{Graduate School of Pure and Applied Sciences, University of Tsukuba, Tsukuba 305-8571, Japan}

\author{Kazuhiro Yabana}
\email[]{yabana@nucl.ph.tsukuba.ac.jp}
\affiliation{Graduate School of Pure and Applied Sciences, University of Tsukuba, Tsukuba 305-8571, Japan}
\affiliation{Center for Computational Sciences, University of Tsukuba, Tsukuba 305-8577, Japan}

\date{May 23, 2016}

\begin{abstract}
\begin{description}
\item[Background]
Multinucleon transfer (MNT) and quasifission (QF) processes are 
dominant processes in low-energy collisions of two heavy nuclei.
They are expected to be useful to produce neutron-rich unstable
nuclei. Nuclear dynamics leading to these processes depends sensitively
on nuclear properties such as deformation and shell structure.

\item[Purpose]
We elucidate reaction mechanisms of MNT and QF processes involving heavy
deformed nuclei, making detailed comparisons between microscopic time-dependent
Hartree-Fock (TDHF) calculations and measurements for the $^{64}$Ni+$^{238}$U reaction.

\item[Methods]
Three-dimensional Skyrme-TDHF calculations are performed.
Particle-number projection method is used to evaluate MNT
cross sections from the TDHF wave function after collision.

\item[Results]
Fragment masses, total kinetic energy (TKE), scattering angle, contact
time, and MNT cross sections are investigated for the $^{64}$Ni+$^{238}$U
reaction. They show reasonable agreements with measurements. At small
impact parameters, collision dynamics depends sensitively on the orientation
of deformed $^{238}$U. In tip (side) collisions, we find a larger (smaller)
TKE and a shorter (longer) contact time. In tip collisions, we find a strong
influence of quantum shells around $^{208}$Pb. 

\item[Conclusions]
It is confirmed that the TDHF calculations reasonably describe both MNT
and QF processes in the $^{64}$Ni+$^{238}$U reaction. Analyses of this
system indicate the significance of the nuclear structure effects such as
deformation and quantum shells in nuclear reaction dynamics at low energies.
\end{description}
\end{abstract}


\maketitle

\section{INTRODUCTION}

It has been known that shell structure and deformation, which are 
fundamental properties of nuclear structure, play an important role 
in low-energy heavy ion reactions. For example, the barrier height for 
nuclear fusion depends on the orientation of colliding nuclei if a deformed
nucleus is involved \cite{Reisdorf(1994),Hagino(2012)}. In the synthesis
of superheavy elements (SHEs), shell effects are crucially important, since
they reduce excitation energy of a compound nucleus, and enhancing its
survival probability. A remarkable example is the successful synthesis of
SHEs by the cold-fusion reactions, where $^{208}$Pb or $^{209}$Bi
target is utilized \cite{Munzenberg(1988),Hofmann(1998)}. Recently,
shell effects on multinucleon transfer (MNT) and quasifission (QF)
processes have also been extensively discussed. These reactions are
expected to be useful to produce neutron-rich unstable nuclei (see,
e.g., Refs.~\cite{Zagrebaev(2008)1,Kozulin(136Xe+208Pb),
Zagrebaev(2013)IQF,Watanabe(2015),Zagrebaev(2014)light_N-rich,
Zagrebaev(2006),Zagrebaev(2008)2,Feng(2009),DNS(2015),Dasso(1994),
Dasso(1995),Mun(2014),Mun(2015)} and references therein).

The present study aims to elucidate reaction mechanisms of
the MNT and QF processes in nuclear reactions involving a heavy
deformed nucleus. Specifically, we focus on $^{64}$Ni+$^{238}$U
reaction for which abundant experimental data of both MNT and QF
processes are available. Since this system has a large $N/Z$
asymmetry [$N/Z=1.29$ ($^{64}$Ni) and 1.59 ($^{238}$U)],
MNT processes toward the charge equilibrium of the total system
are expected. A precise measurement of cross sections of the
MNT processes was carried out by L.~Corradi et~al.~\cite{Corradi(64Ni+238U)}.
Moreover, $^{64}$Ni+$^{238}$U reaction has been expected
as a possible candidate for synthesizing a SHE with $Z=120$.
To examine this possibility, E.M.~Kozulin~et~al.~measured
fragment mass and total kinetic energy (TKE) distributions at
several incident energies \cite{Kozulin(64Ni+238U)}. In
Ref.~\cite{Kozulin(64Ni+238U)}, it was shown that mass-symmetric
fragments are hardly produced in $^{64}$Ni+$^{238}$U reaction.
This fact indicates a strong suppression of the fusion reaction by
QF processes. Mass-angle and mass-TKE distributions including
$^{64}$Ni+$^{238}$U were reported by J.~T{\=o}ke
et~al.~\cite{Toke(QF1985)}.

To investigate MNT and QF processes theoretically, various
models have been developed. For MNT reactions, semiclassical
models called \GRAZING \cite{GRAZING,GRAZING-code,GRAZING-online}
and Complex WKB \cite{CWKB} have been developed with great
successes \cite{Corradi(review)}. The \GRAZING has recently
been extended to incorporate transfer-induced fission, which is
referred to as {\small GRAZING-F} \cite{GRAZING-F}. To describe
damped collisions, a dynamical model based on Langevin-type 
equations \cite{Zagrebaev(2005),Zagrebaev(2007)1,Zagrebaev(2008)1,
Zagrebaev(2013)IQF,Zagrebaev(2014)light_N-rich,Zagrebaev(2006),
Zagrebaev(2008)2,Zagrebaev(2007)2,EXP2015(136Xe+208Pb)},
the dinuclear system (DNS) model \cite{Antonenko(1995),
Adamian(1997)1,Adamian(1997)2,Adamian(1998),Adamian(2003),
Feng(2009),Adamian(2010)1,Adamian(2010)2,Adamian(2010)3,
Mun(2014),Mun(2015),DNS(2015)}, and the improved quantum
molecular dynamics model (ImQMD) \cite{ImQMD(2002),ImQMD(2004),
ImQMD(2008),ImQMD(2013),ImQMD(2015)1,ImQMD(2015)2}
have been extensively developed. Despite numerous successes
in describing measurements, they are to some extent empirical,
containing adjustable parameters. This fact limits their predictive
power. To further extend our understanding of reaction mechanisms
and to improve reliability to predict cross sections, we apply the
microscopic time-dependent Hartree-Fock (TDHF) theory to
MNT and QF processes.

Three-dimensional simulations based on the TDHF theory for
low-energy nuclear reactions started around 70's. They have
been successful to describe various phenomena such as fusion
reactions and deep inelastic collisions \cite{Negele(review),Simenel(review)}.
Applications of the TDHF theory to MNT and QF processes are
rather new. In Ref.~\cite{KS_KY_MNT}, we applied the TDHF
theory to investigate MNT processes in $^{40,48}$Ca+$^{124}$Sn,
$^{40}$Ca+$^{208}$Pb, and $^{58}$Ni+$^{208}$Pb reactions
at energies near the Coulomb barrier, for which precise experimental
data are available \cite{Corradi(40Ca+124Sn),Corradi(48Ca+124Sn),
Corradi(58Ni+208Pb),Szilner(40Ca+208Pb)2}. Applying the particle-number
projection (PNP) method \cite{Projection} to the TDHF wave function
after collision, we evaluated transfer probabilities and cross sections
for each channel specified by the number of neutrons and protons
in the reaction products. From the comparison with measurements,
we showed that the TDHF theory reproduces measured cross sections in
an accuracy comparable to other existing models. In Ref.~\cite{KS_KY_PNP},
we have extended the PNP method to evaluate expectation values of
operators. Recently, we applied our method to an asymmetric system,
$^{18}$O+$^{206}$Pb, at energies above the barrier \cite{Bidyut(2015)}.

Recently, QF processes have been investigated by the TDHF theory
\cite{TDHF(2009)238U-238U,TDHF(2010)IQF,Wakhle(QF2014)Interplay,
UO(QF2014),MyPhD,Hammerton(2015),Washiyama(2015),UO(QF2015)}.
First, QF dynamics in collisions of two actinide nuclei such as $^{238}$U+$^{238}$U
\cite{TDHF(2009)238U-238U} and $^{232}$Th+$^{250}$Cf \cite{TDHF(2010)IQF}
was investigated. In these studies, it has been suggested that QF dynamics
depends sensitively on the nuclear orientations, incident energies, and
impact parameters. In Ref.~\cite{Wakhle(QF2014)Interplay} in which
QF processes of $^{40}$Ca+$^{238}$U were reported, it has been indicated
that shell effects reflecting $Z=82$ and $N=126$ magic numbers have
strong influence in tip collisions, while no shell effect is seen in side collisions.
In that work, it has also been recognized that contact time is much longer
in side collisions than that in tip collisions. The mass-angle distributions (MADs),
which are one of the characteristic observables of QF processes, were
calculated and compared with experimental data, showing reasonable
agreements \cite{Wakhle(QF2014)Interplay,Hammerton(2015)}. 

In this article, we report detailed investigations of MNT and QF
processes in $^{64}$Ni+$^{238}$U reaction performing systematic
TDHF calculations. Since the projectile and the target are open shell nuclei, 
pairing correlations may be important in the collision dynamics. 
However, we ignore the effect of pairing correlation in this study, 
since the inclusion of pairing requires much more computational costs 
which prevent systematic investigations for various initial conditions. 
Making use of the PNP method, we are
able to make detailed comparisons with measurements, including cross
sections. In the studies reported so far \cite{TDHF(2009)238U-238U,
TDHF(2010)IQF,Wakhle(QF2014)Interplay,UO(QF2014),Hammerton(2015),
Washiyama(2015),UO(QF2015)}, sensitive dependence of QF dynamics
on nuclear structure has been suggested. From our detailed analyses
of this system, we expect to elucidate clearly those effects of deformation
and quantum shells on QF processes.

The article is organized as follows: In Sec.~\ref{Sec:method},
we briefly explain the theoretical framework of the TDHF theory
and present computational settings. In Sec.~\ref{Sec:results},
we present the results of our TDHF calculations. In Sec.~\ref{Sec:Elab390},
we investigate $^{64}$Ni+$^{238}$U reaction at $E_{\rm c.m.}\approx 307.35$~MeV.
In Sec.~\ref{Sec:Edep}, incident energy, impact parameter,
and orientation dependence of QF dynamics is investigated.
In Sec.~\ref{Sec:comparison}, we compare the TDHF results
with measurements. In Sec.~\ref{Sec:summary}, summary
and conclusion are presented. A part of the results of the
present analyses was reported in Ref.~\cite{MyPhD}.

\section{METHOD}{\label{Sec:method}}

\subsection{TDHF theory}

We briefly explain our theoretical framework. We start with an action,
\begin{equation}
S = \int_{t_1}^{t_2}
\biggl[\; \sum_{i=1}^A \bigl<\psi_i(t)\big| \zI\hbar\,
\partial_t \big|\psi_i(t)\bigr> - \mathcal{E}[\rho(t)] \,\biggr]\, dt,
\end{equation}
where $\mathcal{E}[\rho(t)]$ denotes an energy density functional
(EDF), which is a functional of various densities and is constructed
so as to reproduce various properties of finite nuclei and nuclear
matter. Applying the stationary condition, $\delta S/\delta \psi_i^* = 0$,
we obtain the TDHF equation,
\begin{equation}
\zI\hbar\,\partial_t\psi_i({\bf r}\sigma q,t)
= \hat{h}[\rho(t)] \psi_i({\bf r}\sigma q,t),
\label{Eq:TDHF}
\end{equation}
where ${\bf r}$ and $\sigma$ are spatial and spin coordinates,
respectively. $q$ ($=n$ or $p$) denotes the isospin of $i$-th nucleon.
Single-particle wave functions, $\psi_i({\bf r}\sigma q,t)$ ($i=1,\cdots,A$),
satisfy the orthonormal relation, $\sum_\sigma\int \psi_i^*({\bf r}\sigma q,t)
\psi_j({\bf r}\sigma q,t)\,d{\bf r}=\delta_{ij}$. Single-particle Hamiltonian,
$\hat{h}[\rho(t)]$, contains a mean-field potential generated by all the
nucleons in the system. The many-body wave function is given by a single
Slater-determinant composed of the single-particle wave functions,
\begin{equation}
\Psi({\bf r}_1\sigma_1q_1,\cdots,{\bf r}_A\sigma_Aq_A,t)
= \frac{1}{\sqrt{A!}}\det\big\{ \psi_i({\bf r}_j\sigma_j q_j,t) \big\}.
\end{equation}
Once the EDF is given, the theory contains no empirical parameters.

In heavy ion reactions, the initial wave function is a Slater determinant
composed of single-particle wave functions of projectile and target nuclei
in their ground state. They are prepared separately by solving the static
Hartree-Fock (HF) equation and are boosted with the relative velocity.
The velocity is evaluated assuming the Rutherford trajectory. For a given
set of incident energy $E$ and impact parameter $b$, the solution of
Eq.~(\ref{Eq:TDHF}) is uniquely determined.

We investigate reactions in which binary reaction products are
produced. To make comparisons with measurements, we analyze
the TDHF wave function at a certain time after collision when the
two fragments are sufficiently separated spatially. We calculate such
quantities as fragment masses, total kinetic energy loss (TKEL),
scattering angle, and MNT cross sections.

\subsection{Computational settings}

We use our own code of TDHF calculations for heavy ion reactions
\cite{MyPhD}. In the code, the TDHF equation is solved in real
space and real time. Single-particle wave functions are represented
on a three-dimensional uniform grid without any symmetry restrictions.
The mesh spacing is set to be 0.8~fm. We employ the 11-point
finite-difference formula for spatial derivatives. The fourth-order
Taylor expansion method is utilized for the time-evolution operator
with a single predictor-corrector step. The time step is set to be
$\Delta t=0.2$~fm/$c$. Hockney's method \cite{Hockney}
is used to calculate the Coulomb potential in the isolated
boundary condition.

A box with $30 \times 30 \times 30$ grid points is used to calculate
the ground state of the projectile and target nuclei. A box with
$70 \times 70 \times 30$ grid points is used for reaction calculations.
We set the incident direction parallel to the $x$-axis and set the impact
parameter vector parallel to the $y$-axis. The initial separation distance
between centers of the projectile and the target is set to be 24~fm along
the incident direction. We stop TDHF calculations when the distance
between centers of the reaction products reaches 26~fm.

\begin{figure}[t]
   \begin{center}
   \includegraphics[width=6.3cm]{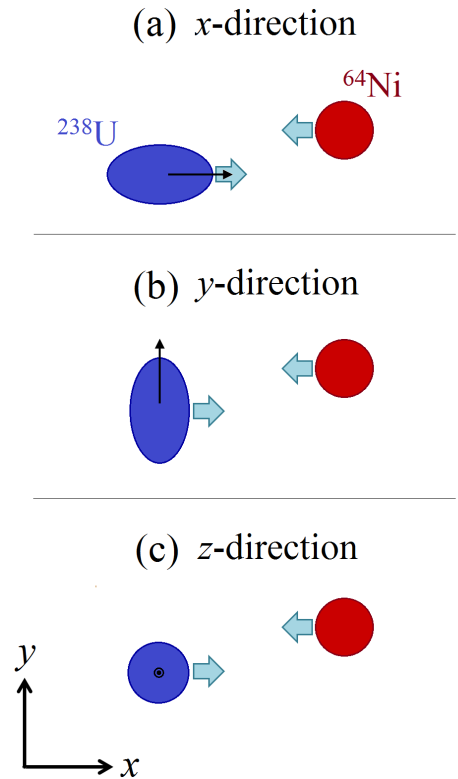}
   \end{center}\vspace{-3mm}
   \caption{(Color online)
   Three initial configurations used for our TDHF calculations of $^{64}$Ni+$^{238}$U
   reaction. (a): The symmetry axis of $^{238}$U is set parallel to the collision
   axis ($x$-axis). (b): The symmetry axis of $^{238}$U is set parallel to the
   impact parameter vector ($y$-axis). (c): The symmetry axis of $^{238}$U
   is set perpendicular to the reaction plane ($xy$-plane).
   }
   \label{FIG:schematic_xyz}
\end{figure}

We use Skyrme SLy5 parameter set \cite{Chabanat} for the EDF.
The ground state of $^{64}$Ni has an oblate shape with $\beta
\approx 0.12$, while that of $^{238}$U has a prolate shape with
$\beta \approx 0.27$. We perform TDHF calculations for three initial
orientations of $^{238}$U: The symmetry axis of $^{238}$U is
set parallel to the incident direction ($x$-axis), set parallel to the
impact parameter vector ($y$-axis), and set perpendicular to the
reaction plane ($xy$-plane). We call these three cases as $x$-,
$y\mbox{-},$ and $z$-direction cases, respectively. Since
deformation of $^{64}$Ni is not very large, we always set the
symmetry axis of $^{64}$Ni perpendicular to the reaction plane,
assuming that the reaction is not affected much by the direction
of the deformed $^{64}$Ni. Figure~\ref{FIG:schematic_xyz}
schematically shows three cases of initial configurations. Since
nuclear rotational motion is very slow, we assume that the nuclear
orientation at the contact of two nuclei can be well specified by
the configurations at the beginning of the TDHF calculations.

\section{TDHF RESULTS}{\label{Sec:results}}

\subsection{Overview of the reaction at E$_{\bf c.m.}\approx$~307.35~MeV}{\label{Sec:Elab390}}

In this Subsection, we show results of TDHF calculations for
$^{64}$Ni+$^{238}$U reaction at $E_{\rm c.m.}\approx 307.35$~MeV.
At around this incident energy, several measurements have been
reported \cite{Corradi(64Ni+238U),Kozulin(64Ni+238U),Toke(QF1985)}.
Comparisons of the TDHF results with the measurements will be presented
in Sec.~\ref{Sec:comparison}. The calculations are performed for an
impact parameter range, 0~fm~$\le b \le 12$~fm. We evaluate the
frozen HF barrier as described in Ref.~\cite{KS_KY_MNT}. The barrier
height is evaluated to be 242.93~MeV for $x$-direction case and 263.97~MeV
for $y$-direction case. The incident center-of-mass energy of $E_{\rm c.m.}
\approx 307.35$~MeV corresponds to about 27~\% and 16~\% above the
barrier for the $x$- and $y$-direction cases, respectively, at $b=0$~fm.
In our TDHF calculations at this incident energy, we always found binary
reaction products and no fusion reaction was observed even in head-on
collisions.

We will show scattering angle in the center-of-mass frame, $\theta_
{\rm c.m.}$, TKEL, and contact time. The scattering angle and the
TKEL are evaluated from the translational motion of reaction products
as described in Ref.~\cite{KS_KY_MNT}. The contact time is defined
as the duration in which the lowest density between colliding nuclei
exceeds a half of the nuclear saturation density, $\rho_0/2=0.08$~fm$^{-3}$.
The same definition was also used by other authors \cite{Wakhle(QF2014)Interplay}.

\begin{figure}[t]
   \begin{center}
   \includegraphics[width=6.6cm]{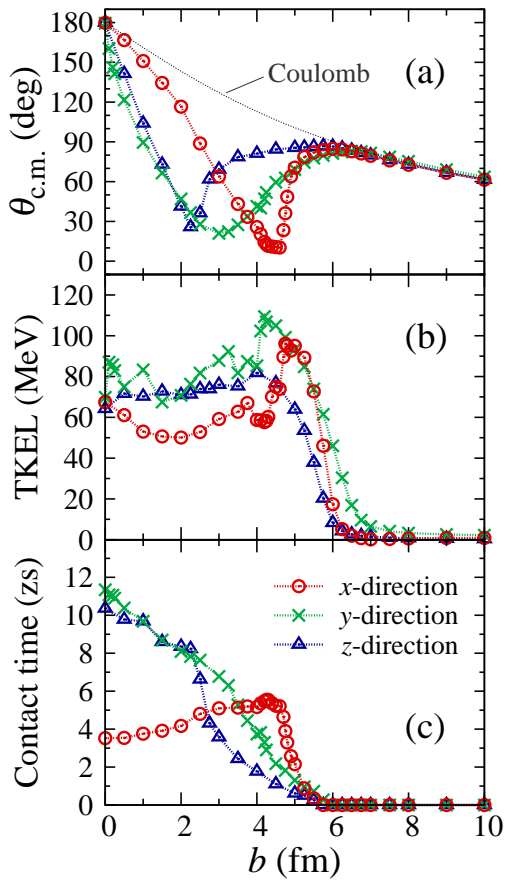}
   \end{center}\vspace{-3mm}
   \caption{(Color online)
   TDHF results for $^{64}$Ni+$^{238}$U reaction at $E_{\rm c.m.}\approx 307.35$~MeV.
   Scattering angle in the center-of-mass frame, $\theta_{\rm c.m.}$, total kinetic
   energy loss (TKEL), and contact time are shown in (a), (b), and (c), respectively,
   as functions of the impact parameter, $b$. Results for $x$-, $y$-, and $z$-direction
   cases are shown by red circles, green crosses, and blue open triangles connected
   with dotted lines, respectively. In (a), the scattering angle for the Rutherford
   trajectory is shown by a dotted curve.
   }
   \label{FIG:theta+TKEL_vs_b}
\end{figure}

\begin{figure}[t]
   \begin{center}
   \includegraphics[width=8.6cm]{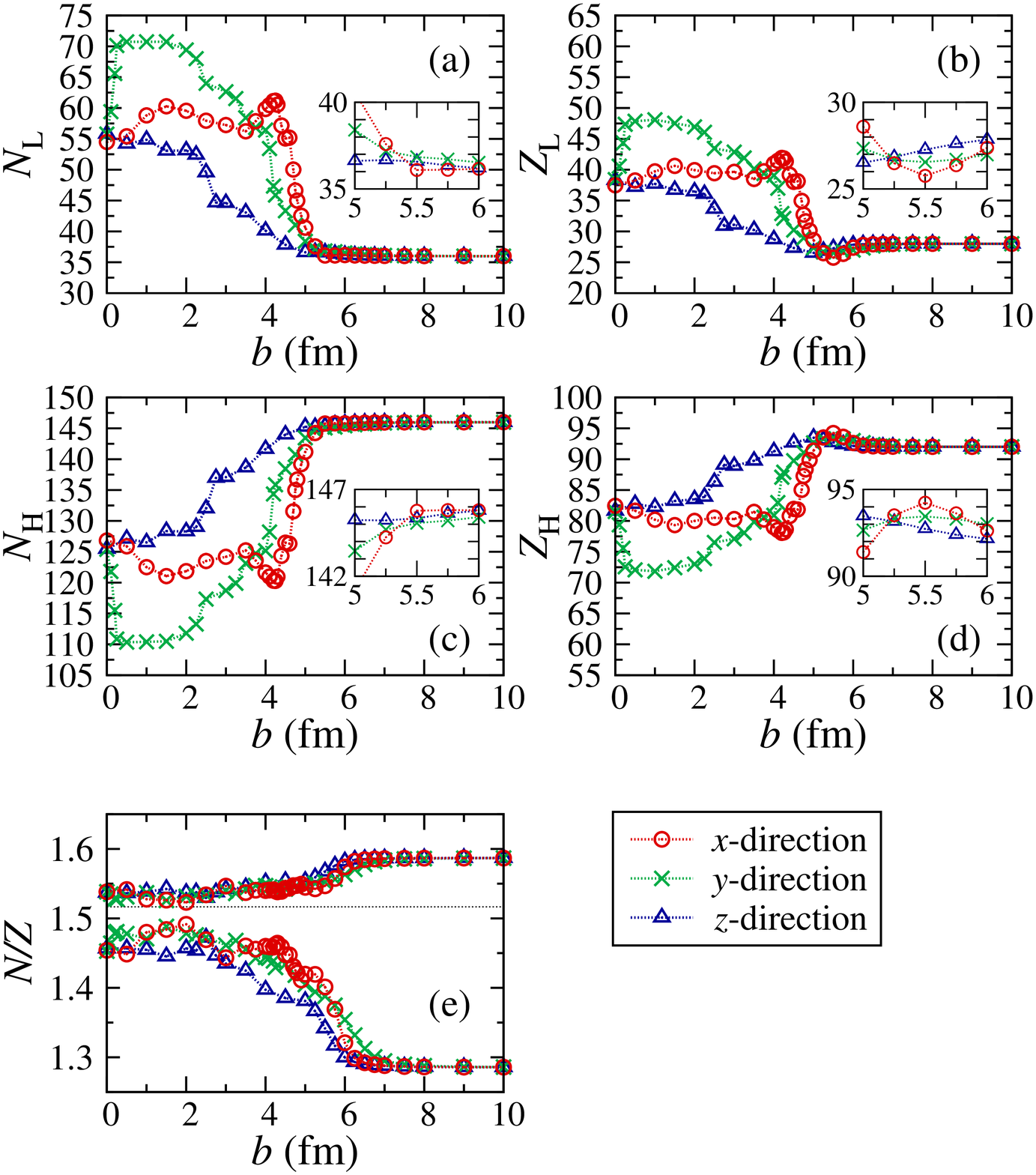}
   \end{center}\vspace{-3mm}
   \caption{(Color online)
   TDHF results for $^{64}$Ni+$^{238}$U reaction at $E_{\rm c.m.}\approx 307.35$~MeV.
   Average numbers of nucleons in lighter (a,~b) and heavier (c,~d) fragments
   are shown as functions of the impact parameter, $b$. Left panels (a,~c) show those
   of neutrons, while right panels (b,~d) show those of protons. Insets are magnified
   plots of an impact parameter region, $b=5\mbox{--}6$~fm. The neutron-to-proton
   ratios of lighter and heavier fragments are shown in (e). Results for $x$-, $y$-,
   and $z$-direction cases are shown by red circles, green crosses, and blue open
   triangles connected with dotted lines, respectively. In (e), the fully equilibrated
   value of the system, 1.52, is indicated by a horizontal dotted line.
   }
   \label{FIG:Nave}
\end{figure}

In Fig.~\ref{FIG:theta+TKEL_vs_b}, we show $\theta_{\rm c.m.}$,
TKEL, and contact time in (a), (b), and (c), respectively, as functions of 
the impact parameter. Results for $x$-, $y$-, and $z$-direction cases are
shown by red circles, green crosses, and blue open triangles connected
with dotted lines, respectively. The same symbols will be used in
Figs.~\ref{FIG:Nave}, \ref{FIG:Wilczynski}, \ref{FIG:MAD}, and \ref{FIG:TKE-A}.
In (a), the scattering angle for the Rutherford trajectory is shown by a dotted
curve. In (c), contact time is shown in zeptosecond ($1$~zs~$=10^{-21}$~sec).

We first investigate behavior which does not depend much on the initial
orientation of $^{238}$U. When the impact parameter is sufficiently large
($b \gtrsim 7$~fm), the reaction is governed by the Coulomb interaction,
and the scattering angle coincides with that of the Rutherford trajectory.
At this impact parameter region, TKEL is very small and contact time is
zero. As the impact parameter decreases ($b \lesssim$~7~fm), TKEL
increases rapidly taking maximum values at $b \approx 4\mbox{--}5$~fm.
Surfaces of two nuclei also start to touch gently, and the nuclear attractive 
interaction distorts trajectory toward forward angles.

At a small-$b$ region ($b \lesssim 5$~fm), the contact time becomes
substantially long. This indicates a formation of a dinuclear system
connected by a thick neck. Because the dinuclear system rotates for a 
certain period, the scattering angle decreases noticeably as shown in (a). 
As the impact parameter decreases further, the scattering angle increases 
monotonically, reaching $180^\circ$ (backward scattering) in head-on
collisions. In this small-$b$ region, TKEL is roughly constant.

We next look at dependence on the orientation of $^{238}$U
seen in Fig.~\ref{FIG:theta+TKEL_vs_b}. As the impact parameter
decreases from $b\approx 6$~fm, we observe a rapid decrease of the
scattering angle in (a). In contrast, we observe a rapid increase of the
contact time in (c). The decrease (increase) of the scattering angle
(contact time) is steepest for the $x$-direction case and becomes
moderate as the orientation changes from $x$- to $y$- and from
$y$- to $z$-direction. This difference can be understood as follows.
In the $x$-direction case, the symmetry axis of $^{238}$U is set
parallel to the collision axis (Fig.~\ref{FIG:schematic_xyz}~(a)).
In this geometry, two nuclei collide substantially at a large impact
parameter, $b \approx 5$~fm, compared to the other cases. In the
$z$-direction case, $^{64}$Ni always collides with the side of
$^{238}$U  (Fig.~\ref{FIG:schematic_xyz}~(c)). This results
in the slowest change of $\theta_{\rm c.m.}$ and contact time.
Results of the $y$-direction case (Fig.~\ref{FIG:schematic_xyz}~(b))
locate between those of the $x$- and $z$-direction cases.

The contact time shown in (c) has a strong orientation
dependence at a small-$b$ region ($b \lesssim 4$~fm).
In the $y$- and $z$-direction cases, contact time increases
monotonically as the impact parameter decreases, reaching
10--11~zs in head-on collisions. On the other hand, in the
$x$-direction case, contact time takes almost a constant
value (about 4--5~zs), even decreases slightly as the impact
parameter decreases. Because of the shorter contact time,
the combined dinuclear system does not rotate much. This
explains larger scattering angles for the $x$-direction case
compared with the other cases at small impact parameters
($b \lesssim 3$~fm), seen in (a). The observed orientation 
dependence of the contact time is consistent with the TDHF 
calculations for $^{40}$Ca+$^{238}$U reported in
\cite{Wakhle(QF2014)Interplay}.

\begin{figure*}[t]
   \begin{center}
   \includegraphics[width=17.7cm]{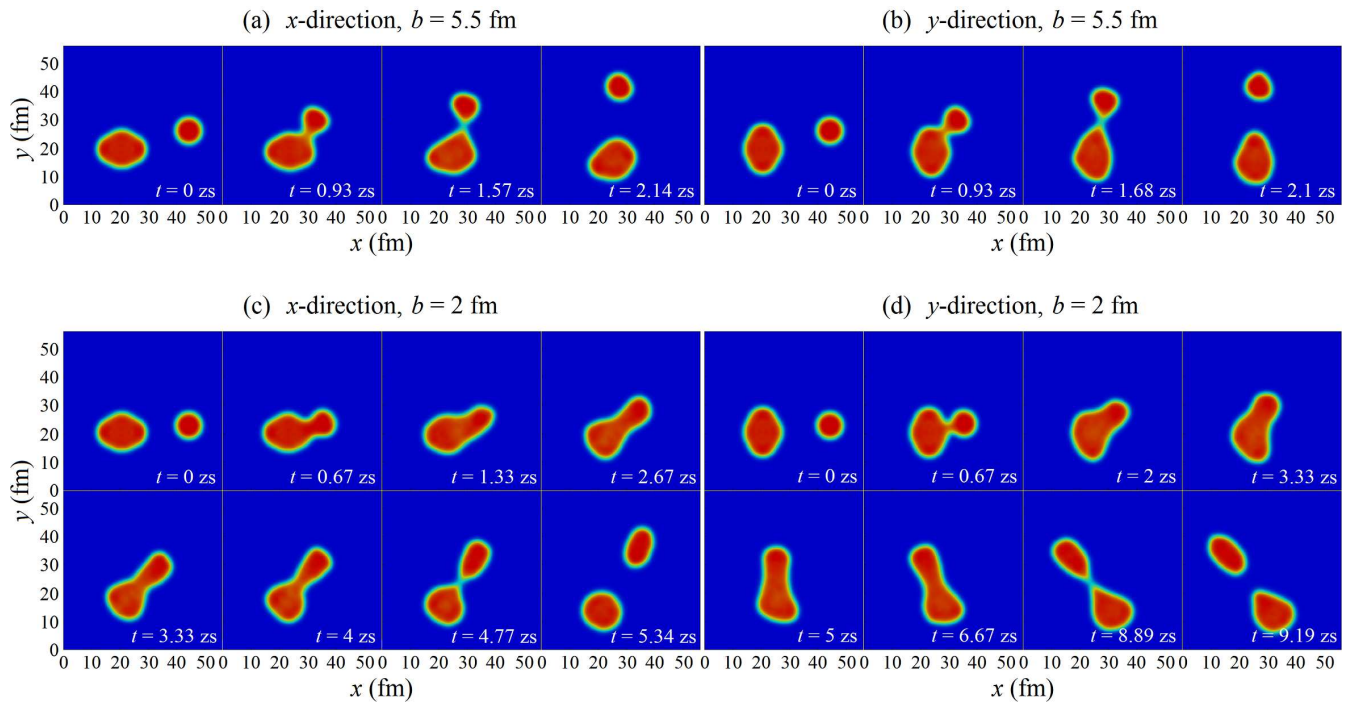}
   \end{center}\vspace{-3mm}
   \caption{(Color online)
   Snapshots of the density in the reaction plane in TDHF calculations for
   $^{64}$Ni+$^{238}$U reaction at $E_{\rm c.m.}\approx 307.35$~MeV.
   Results for two impact parameters, $b=0.5$~fm and 2~fm, and two
   initial orientations of $^{238}$U, the $x$- and $y$-direction cases,
   are shown. Elapsed time measured from the initial configuration is
   indicated in each panel in zeptosecond ($1$~zs~$=10^{-21}$~sec).
   See also Supplemental Material \cite{SM} for movies of the reactions.
   }
   \label{FIG:rho(t)}
\end{figure*}

To obtain intuitive understanding of the reaction dynamics,
we show in Fig.~\ref{FIG:rho(t)}~(a-d) snapshots of the density
in the reaction plane for two impact parameters, 5.5~fm and 2~fm,
and two orientations of $^{238}$U, the $x$- and $y$-direction cases.
Elapsed time measured from the initial configuration is indicated in
zeptosecond. At $b=5.5$~fm shown in (a,~b), we find a formation
of a thin neck through which a few nucleons are exchanged. The
reaction dynamics does not show much difference between the $x$-
and $y$-direction cases at this impact parameter.

Contrarily, we find quite different reaction dynamics at a small-$b$
reaction, $b=2$~fm, for different orientations of $^{238}$U. Let us
first look at reaction dynamics in the $x$-direction case at
$b=2$~fm shown in (c). As time evolves, $^{64}$Ni collides with
$^{238}$U at a position close to the tip of the $^{238}$U ($t=0.67$~zs).
Then a thick and long neck is developed in the course of the collision,
forming an elongated dinuclear system ($t=0.67$--2.67~zs). After
this stage, the neck becomes thinner ($t=3.33$--4~zs) and eventually
ruptures ($t=4.77$~zs), producing binary reaction products ($t=5.34$~zs).
The produced fragments roughly correspond to $_{40}^{100}$Zr$_{60}^{}$
and $_{80}^{202}$Hg$_{122}^{}$. We note that we have found very similar
shape evolution dynamics to that shown in Fig.~\ref{FIG:rho(t)}~(c) in
a wide impact parameter range of $b=0 \mbox{--}4$~fm, where the contact
time is almost constant as shown in Fig.~\ref{FIG:theta+TKEL_vs_b}~(c)
(see also Supplemental Material \cite{SM} for movies of the reactions).

Figure~\ref{FIG:rho(t)}~(d) shows reaction dynamics in the $y$-direction
case at $b=2$~fm. In this case, $^{64}$Ni collides with $^{238}$U at
a position close to the side of the $^{238}$U ($t=0.67$~zs). After the
touch, a somewhat compact composite system with a thick neck structure
is formed ($t=2$~zs) (Note that the time of each snapshot is not the same
as that shown in (c)). The dinuclear system with a thick neck structure is
maintained for a long period and rotates in the reaction plane ($t=2
\mbox{--}6.67$~zs). When the neck ruptures ($t=8.89$~zs), fragments
with more symmetric masses are generated compared with those of the
$x$-direction case shown in (c). The produced fragments roughly correspond
to $_{47}^{116}$Ag$_{69}^{}$ and $_{73}^{185}$Ta$_{112}^{}$.

We next investigate average numbers of nucleons in the reaction
products as functions of the impact parameter. Figure~\ref{FIG:Nave}~(a)
and (b) show average numbers of neutrons and protons in the lighter
fragment, which we denote as $N_{\rm L}$ and $Z_{\rm L}$, respectively.
Those in the heavier fragment, which we denote as $N_{\rm H}$ and
$Z_{\rm H}$, are shown in (c) and (d), respectively. $N/Z$ ratios of
the lighter and the heavier fragments are also shown in (e). In (e),
the fully equilibrated value of the system, 1.52, is indicated by
a horizontal dotted line. 

When the impact parameter is sufficiently large ($b \gtrsim7$~fm),
the average numbers of neutrons and protons coincide with those
of the projectile and target nuclei. As the impact parameter decreases
($b\approx 5\mbox{--}6$~fm), we find that protons are transferred
from $^{64}$Ni to $^{238}$U, while neutrons tend to be transferred
in the opposite direction. These directions of transfers correspond
to those expected form the initial $N/Z$ asymmetry. We show a
magnified plot for this impact parameter range as insets in (a-d).
The snapshots of the density shown in Fig.~\ref{FIG:rho(t)}~(a,~b)
correspond to reactions in this impact parameter range.

As the impact parameter decreases further, we find a drastic change
at around $b\approx 4\mbox{--}5$~fm. Inside this impact parameter,
a mass equilibration process toward the direction increasing the
mass symmetry, which we call the mass-drift mode, is observed.
In the mass-drift mode, both neutrons and protons are transferred
toward the same direction, from the heavier nucleus to the lighter one.
While the fragment masses show substantial changes at $b\approx 4\mbox{--}5$~fm,
the $N/Z$ ratios approach monotonically to the fully equilibrated value.
From the density profile during the reaction, we find that the shape evolution 
and the neck rupture are responsible for the mass-drift mode. Once
a dinuclear system is formed in the course of collision, the system
quickly reaches the charge equilibrium, and the position of the neck
rupture determines the amount of transfers of neutrons and protons.
In Ref.~\cite{KS_KY_MNT}, we reported similar transfer dynamics
in lighter systems.

The mass-drift mode observed at $b \lesssim 5$~fm shows noticeable
dependence on the initial orientation. In the $z$-direction case (blue open
triangles), we find a gradual change of the average number of nucleons.
In contrast, in the $x$- and $y$-direction cases, we observe an abrupt
change at $b \approx 4\mbox{--}5$~fm. In the $x$-direction case
(red open circles), the average number of nucleons exhibits a prominent
plateau which persists within 0~fm~$\le b \lesssim 4$~fm. In this
impact parameter region, $N_{\rm H} \approx 120\mbox{--}126$ and
$Z_{\rm H}\approx 78\mbox{--}82$ are observed. We consider that
the quantum shells of $^{208}$Pb make a significant contribution
to this behavior. A similar shell effect of $^{208}$Pb has been
reported in the tip collisions of $^{40}$Ca+$^{238}$U in TDHF
calculations \cite{Wakhle(QF2014)Interplay}. We note that, in our
calculations, the lighter partner has $N_{\rm L}\approx 55\mbox{--}60$
and $Z_{\rm L}\approx 37\mbox{--}42$. A production of similar fragments
has been observed in TDHF calculations for the side collisions
of $^{40,48}$Ca+$^{238}$U at $b=0$~fm \cite{UO(QF2014)}.
A possible influence of stabilization by strongly bound Zr isotopes
with large prolate deformation in this mass region has been
advocated \cite{UO(QF2014)}.

In the $y$-direction case (green crosses), the behavior is quite
different. As in the $x$-direction case, we observe an abrupt change
of the average number of nucleons at $b \approx 4\mbox{--}5$~fm.
However, the plateau around $N_{\rm H}\approx 126$ and $Z_{\rm H}
\approx 82$ does not appear. The composite system tends to split into
more mass-symmetric fragments. It indicates that the quantum shells
of $^{208}$Pb are not significant in this case. A similar interplay
between the quantum shells and the nuclear orientation was reported
in $^{40}$Ca+$^{238}$U \cite{Wakhle(QF2014)Interplay}.
In Ref.~\cite{Wakhle(QF2014)Interplay}, it was reported that quantum
shells do not contribute in the side collisions of $^{40}$Ca+$^{238}$U.
Contrarily to it, we find another plateau behavior in the $y$- and
$z$-direction cases. In the $y$-direction case, at a small-$b$ region,
0~fm~$\lesssim b \lesssim 2$~fm, we observe a plateau at around
$N_{\rm H}\approx 110$ and $Z_{\rm H}\approx 72$ for the heavier
fragment and $N_{\rm L}\approx 70$ and $Z_{\rm L}\approx 48$ for
the lighter fragment. This behavior may be influenced by the shell
effect of $Z=50$ in the QF process, although the fragment shows
a large deformation as shown in Fig.~\ref{FIG:rho(t)}~(d). In the $z$-direction
case, a plateau is seen at around $N_{\rm H}\approx 127$ and $Z_{\rm H}
\approx 83$ for the heavier fragment and $N_{\rm L}\approx 54$ and
$Z_{\rm L}\approx 37$ for the lighter fragment. This behavior indicates
the effect of the quantum shells of $^{208}$Pb. We note that
influence of quantum shells in QF processes has been routinely
observed experimentally \cite{Itkis(2004),Nishio(36S+238U),
Nishio(34S+238U),Nishio(4048Ca+238U),Kozulin(2014)1,
Kozulin(2014)IOP,Kozulin(2014)2} and discussed theoretically
\cite{Zagrebaev(2005),Zagrebaev(2007)1,Zagrebaev(2013)IQF,
Zagrebaev(2006),Zagrebaev(2007)2,Wakhle(QF2014)Interplay,
Aritomo(2004),Aritomo(2005),Aritomo(2009),Aritomo(2012)}.

It is worth emphasizing that, in the $y$-direction case, the average
number of nucleons changes dramatically when the impact parameter
becomes a tiny but a finite value. For instance, from $b=0$ to 0.25~fm,
the average number of nucleons changes as large as 25. We consider
that the observed behavior is related to the symmetry that appears
only at $b=0$ fm in which the colliding system has a rotational symmetry
around the collision axis. This symmetry disappears once the impact
parameter becomes finite.

We note that the behavior at around $b=0$ fm is different between
$y$- and $z$-direction cases. To understand the origin of the difference,
let us consider shape of the system viewed from a frame rotating with the
vector connecting centers of the two colliding nuclei, ${\bf R}(t)$, in the
adiabatic limit neglecting currents. In the $z$-direction case, the system always
persists a reflection symmetry with respect to the plane which contains
${\bf R}(t)$ and is perpendicular to the reaction plane. On the other hand,
in the $y$-direction case at a nonzero impact parameter, the system does not
have the symmetry mentioned above due to the deformed shape of $^{238}$U.
Thus the system may go through more complex shape evolution dynamics.
In fact, once the impact parameter becomes nonzero in the $y$-direction case,
we find the projectile-like subsystem moving along the elongated direction of
the $^{238}$U-like subsystem forming a very thick neck, which results in the
abrupt change of the average number of transferred nucleons (See
Supplemental Material \cite{SM} for movies of the reactions).

The orientation dependence is also clearly seen in the TKEL at
a small-$b$ region ($b \lesssim 4$~fm) in Fig.~\ref{FIG:theta+TKEL_vs_b}~(b).
In the $y$- and $z$-direction cases, TKEL takes almost constant
values, $\approx 70\mbox{--}80$~MeV. We observe somewhat larger
values of TKEL in the $y$-direction case compared with those of the
$z$-direction case. This difference may reflect the reflection symmetry
mentioned above which restricts reaction dynamics in the $z$-direction case.
In the $x$-direction case, we observe smaller values, $\approx 50\mbox{--}60$~MeV.
In Ref.~\cite{Guillaume(2015)258Fm}, fission dynamics of $^{258}$Fm
was investigated by TDHF+BCS approach. It was shown that the TKE
exhibits clear dependence on the shape of the fissioning nucleus,
and that the different shape evolution dynamics is associated with
different valleys in the potential energy surface (PES). Although
we have not conducted PES calculations of $_{120}^{302}$Ubn$_{182}^{}$
composite system, we expect that there exists a valley in the PES
of the composite system associated with the doubly magic $^{208}$Pb
and that the valley causes the small TKEL and the short contact time.

We note that an experimentally measured TKE distribution of 
$^{64}$Ni+$^{238}$U reaction at a smaller incident energy, 
$E_{\rm c.m.}\approx 282.13$~MeV, was reported \cite{Kozulin(64Ni+238U)}. 
In the measurement, a two-peaked structure of TKE was observed. 
Although the plot was constructed from selected fragments having
$A_{\rm CN}/2 \pm 20$, it is expected that different dynamics 
associated with the large deformation of $^{238}$U affects the
measured trends.

\subsection{Incident energy dependence}{\label{Sec:Edep}}

In this Subsection, we examine incident energy dependence of QF
processes in $^{64}$Ni+$^{238}$U reaction. We investigate reactions
at two impact parameters, $b=0.5$~fm and 2~fm, for two orientations
of $^{238}$U, $x$- and $y$-direction cases.

\begin{figure}[t]
   \begin{center}
   \includegraphics[width=8.6cm]{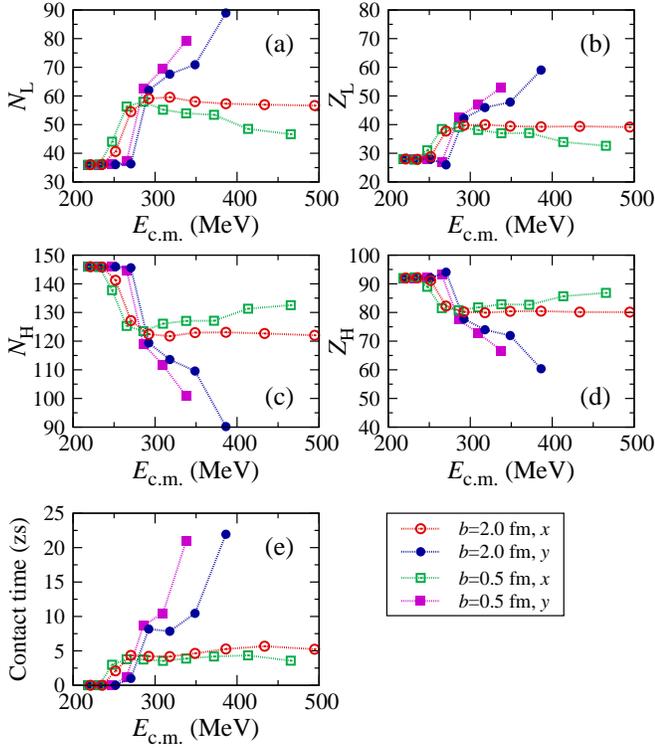}
   \end{center}\vspace{-3mm}
   \caption{(Color online)
   TDHF results for $^{64}$Ni+$^{238}$U reaction at $b=0.5$~fm and
   2~fm for $x$- and $y$-direction cases. Average numbers of nucleons in
   lighter (a,~b) and heavier (c,~d) fragments are shown as functions
   of the center-of-mass energy, $E_{\rm c.m.}$. Left panels (a,~c) show
   those of neutrons, while right panels (b,~d) show those of protons.
   In (e), contact time is presented. Results for $b=0.5$- and $2$-fm
   cases are shown by squares and circles, and those for $x$- and
   $y$-direction cases are shown by open and filled symbols, respectively.
   }
   \label{FIG:Nave_vs_E}
\end{figure}

Figure~\ref{FIG:Nave_vs_E}~(a) and (b) show average numbers of
neutrons and protons in the lighter fragment, respectively. Those in the
heavier fragment are shown in (c) and (d). In (e), contact time is also
presented. The horizontal axis is the center-of-mass energy, $E_{\rm c.m.}$.

First, we consider the $x$-direction case (open symbols). As
the center-of-mass energy increases, we find an abrupt change
in the fragment masses when the energy exceeds the barrier height,
$V_{\rm B}\approx 242.93$~MeV. Just above the barrier, the fragment
masses are about $N_{\rm L}\approx 58$ and $Z_{\rm L}\approx 40$
and $N_{\rm H}\approx 124$ and $Z_{\rm H}\approx 80$ for both
$b=0.5$- and 2-fm cases.

For $b=2$~fm case (red open circles), the fragment masses are
almost independent of the center-of-mass energy for an energy
range, 290~MeV~$\lesssim E_{\rm c.m.}\lesssim$~500~MeV.
It indicates a significant influence of the quantum shells of
$^{208}$Pb, even above barrier energies. On the other hand,
for $b=0.5$~fm case (green open squares), the amount of
transferred nucleons decreases as the center-of-mass energy
increases. This behavior implies that the shell effect is weakened
as the incident energy increases for $b=0.5$~fm case. In the
$x$-direction case, an elongated dinuclear system is observed
even at energies well above the barrier (See also Supplemental
Material \cite{SM}). Because of the large elongation, the dinuclear
system splits in a relatively short period ($\approx 4$--5~zs) as seen
in Fig.~\ref{FIG:Nave_vs_E}~(e), and no fusion reaction was
observed for all incident energies examined here.

Next, we consider the $y$-direction case (filled symbols).
For both $b=0.5$- and 2-fm cases, we observe similar behavior
as a function of the center-of-mass energy. As in the $x$-direction
case, we find an abrupt change in the fragment masses when the
center-of-mass energy exceeds the barrier height, $V_{\rm B}\approx
263.97$~MeV. In contrast to the $x$-direction case, the fragment
masses continue to change as the center-of-mass energy increases, 
up to $E_{\rm c.m.}\approx 338$ (386)~MeV for $b=0.5$ (2)~fm.
We also find an abrupt change in the contact time in (e). In the
$y$-direction case, the composite system shows a compact shape,
which becomes a mononuclear shape as the center-of-mass energy
increases. The mononuclear system splits into mass-symmetric fragments.
As a result of the mononuclear system formation, the contact time
becomes much longer in the $y$-direction case than that in the
$x$-direction case, as shown in (e).

We note that, in the $y$-direction case at higher center-of-mass
energies, $E_{\rm c.m.}\gtrsim 338$ (386)~MeV for $b=0.5$ (2)~fm,
a capture process takes place, forming a superheavy composite system
with $Z=120$. We continued time-evolution calculations up to $40$~zs
(60,000 time steps). Similar criteria for fusion were also used by other
authors \cite{Wakhle(QF2014)Interplay,UO(QF2014)}. In this period,
the composite system exhibits a compact mononuclear shape (See
also Supplemental Material \cite{SM}). In Ref.~\cite{Kozulin(64Ni+238U)},
measured fragment mass distributions in $^{64}$Ni+$^{238}$U reaction
were reported at several incident energies. They showed that mass-symmetric
fragments are hardly produced in the reaction. Our results are consistent
with the experimental observation, since the highest incident energy of
the experiment was $E_{\rm c.m.}\approx 301.05$~MeV, and is much
smaller than the present threshold energy for fusion in our TDHF calculations.
Our results indicate that more mass-symmetric fragments will be produced
after forming a mononuclear system at higher incident energies, although
it should accompany substantial excitation energy. We note that recent
experimental data \cite{Z120(2008),Z120(2012)} show that the superheavy
element with $Z=120$ could be formed by $^{64}$Ni+$^{238}$U reaction
at $E_{\rm c.m.}\approx 332.88$~MeV, which lived longer than $10^{-18}$~s.

\section{Comparison with measurements}{\label{Sec:comparison}}

\subsection{Production cross sections}\label{Sec:NTCS}

\begin{figure}[t]
   \begin{center}
   \includegraphics[width=8.6cm]{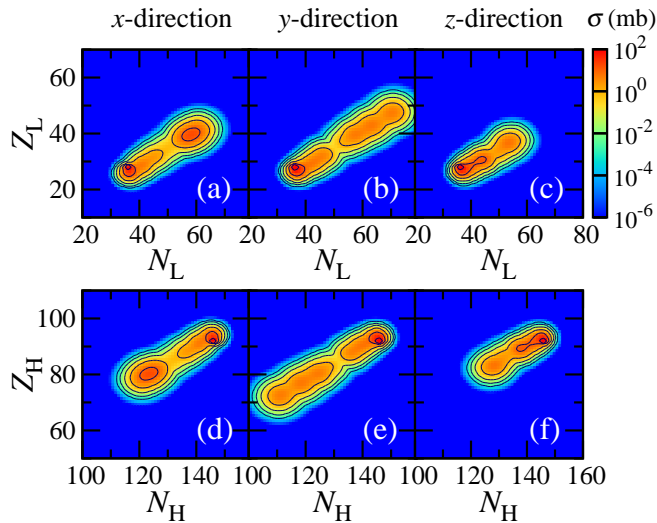}
   \end{center}\vspace{-4mm}
   \caption{(Color online)
   Primary production cross sections, $\sigma(N,Z)$, for $^{64}$Ni+$^{238}$U
   reaction at $E_{\rm c.m.}\approx 307.35$~MeV in TDHF calculation. Upper panels
   (a-c) show cross sections for lighter fragments, while lower panels (d-f)
   show those for heavier fragments. Contributions from $x$-, $y$-, and
   $z$-direction cases are shown in left, middle, and right panels, respectively.
   The contour lines correspond to $\sigma=100$, 10, 1, 0.1, and 0.01~mb.
   }
   \label{FIG:sigma_NZ}
\end{figure}

To compare with measured cross sections of MNT processes,
we employ the PNP method \cite{Projection,KS_KY_MNT,KS_KY_PNP}.
We use the PNP operator,
\begin{equation}
\hat{P}_n^{(q)} = \frac{1}{2\pi} \int_0^{2\pi}
e^{\zI (n-\hat{N}_V^{(q)})\theta}d\theta,
\end{equation}
where $\hat{N}_V^{(q)}$ is the number operator in a volume $V$.
The probability that $n$ nucleons are included in $V$ is given by
\begin{equation}
P_n^{(q)} = \frac{1}{2\pi} \int_0^{2\pi} e^{\zI n\theta}
\det\mathcal{B}^{(q)}(\theta)\, d\theta,
\label{Eq:P_n}
\end{equation}
where
\begin{equation}
\mathcal{B}_{ij}^{(q)}(\theta) =
\sum_\sigma \int \psi_i^*({\bf r}\sigma q)\psi_j({\bf r}\sigma q)
\bigl( \Theta_{\bar{V}}({\bf r}) + e^{-\zI \theta}\Theta_V({\bf r}) \bigr) d{\bf r}.
\end{equation}
$\Theta_{V\,(\bar{V})}({\bf r})$ denotes a space division function
which is equal to 1 inside $V$ ($\bar{V}$) and 0 elsewhere. $\bar{V}$
is the complement of $V$. In practice, the integral in Eq.~(\ref{Eq:P_n})
is evaluated using the trapezoidal rule discretizing the interval into
300 equal grids. The production cross section for a reaction product
composed of $N$ neutrons and $Z$ protons is given by
\begin{equation}
\sigma(N,Z) = 2\pi \int_0^\infty b\: P_{N,Z}(b)\, db,
\label{Eq:sigmatot}
\end{equation}
where $P_{N,Z}$ takes a product from, $P_N^{(n)} P_Z^{(p)}$, in the TDHF theory.

\begin{figure*}[t]
   \begin{center}
   \includegraphics[width=16.5cm]{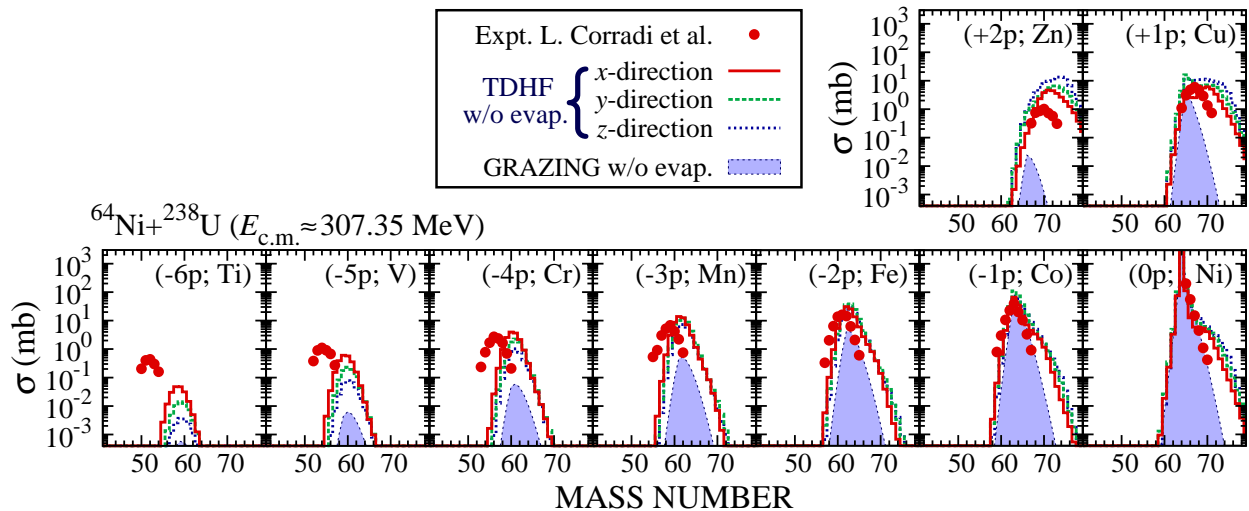}
   \end{center}\vspace{-3mm}
   \caption{(Color online)
   Transfer cross sections for $^{64}$Ni+$^{238}$U reaction at $E_{\rm c.m.}\approx 307.35$~MeV.
   Each panel shows cross sections for different proton-transfer channel indicated by
   ($\pm x$p;~X), where X stands for the corresponding element. The horizontal axis
   is the mass number of lighter fragments. Experimental data \cite{Corradi(64Ni+238U)}
   are shown by red filled circles. TDHF results in $x$-, $y$-, and $z$-direction cases
   are represented by red solid, green dashed, and blue dotted histograms, respectively.
   We also show cross sections calculated by the \GRAZING code \cite{GRAZING-online} with
   standard parameter sets $^{\mbox{\scriptsize\ref{FootNote1}}}$.
   }
   \label{FIG:NTCS}
\end{figure*}

In Fig.~\ref{FIG:sigma_NZ}, we show production cross sections,
$\sigma(N,Z)$, for $^{64}$Ni+$^{238}$U reaction at $E_{\rm c.m.}
\approx 307.35$~MeV. Upper panels (a-c) show cross sections for lighter fragments,
while lower panels (d-f) show those for heavier fragments. We show
cross sections for $x$-, $y$-, and $z$-direction cases in left,
middle, and right panels, respectively. To compare with measurements,
we should take a proper average over the orientations of $^{238}$U. 
We did not do it, since it requires too much computational costs.

From the figure, we find that the cross sections extend
widely in the $N$-$Z$ plane. There is a peak of $\sigma(N,Z)$ 
at around $(N_{\rm L},Z_{\rm L})=(36, 28)$ in (a-c) for
lighter fragments and $(N_{\rm H}, Z_{\rm H})=(146, 92)$ in (d-f)
for heavier fragments. They are contributed from a large-$b$
region, $b \gtrsim 5$~fm. We also find a peak in $\sigma(N,Z)$
located inside a region of $N_{\rm L}>50$, $Z_{\rm L}>30$ in
(a-c) and $N_{\rm H}<130$, $Z_{\rm H}<90$ in (d-f). They are
produced by the QF processes accompanying a large mass-drift
toward the mass symmetry, which take place in a small-$b$ region,
$b \lesssim 4$~fm. The appearance of separated peaks in the
$N$-$Z$ plane is caused by the abrupt change of the reaction
mechanism from quasielastic and MNT to QF at $b\approx
4\mbox{--}5$~fm. The peak positions are consistent with
the observation in Fig.~\ref{FIG:Nave}.

In Ref.~\cite{Corradi(64Ni+238U)}, experimentally measured
transfer cross sections for $^{64}$Ni+$^{238}$U reaction at
$E_{\rm c.m.}\approx 307.35$~MeV were reported. In Fig.~\ref{FIG:NTCS},
we show a comparison of transfer cross sections between our TDHF
results and the measurements as a function of the mass number
of lighter fragments. Each panel shows cross sections for different
proton-transfer channel. The number of transferred protons is
indicated by ($\pm x$p;~X), where X stands for the corresponding
element. The plus sign is for transfer processes from $^{238}$U
to $^{64}$Ni (pickup), while the minus sign is for the opposite direction
(stripping). Measured cross sections are shown by red filled circles.
TDHF results in $x$-, $y$-, and $z$-direction cases are shown
by red solid, green dashed, and blue dotted histograms,
respectively. Cross sections calculated by the \GRAZING code
\cite{GRAZING-online} using standard input parameters
\footnote{\label{FootNote1}
Input parameters that we used for the \GRAZING calculation:
For low-lying excitations: $E_2=1.35$ (0.04)~MeV, $B(E_2)=0.076$ (12.09)~$e^2$b$^2$,
$E_3=3.56$ (0.73)~MeV, $B(E_3)=0.022$ (0.58)~$e^2$b$^2$; for giant resonances:
$E_2=57$~(94)~$A^{-1/3}$~MeV, strength~=~0.8 (0.4)~\% of sum rule, width~=~2.5 (6);
for single-particle states: $\delta^\nu=8$, $\delta^\pi=8$, level density~=~2.455 (2.053) MeV$^{-1}$
(neutron), 10.527 (8.298)~MeV$^{-1}$ (proton); these values are for projectile (target).
}
are also shown by filled areas.

We note that experimental data are suffered by effects of particle
evaporation from excited reaction products, whereas the TDHF results
correspond to primary cross sections just after the reseparation.
In addition, the measurement was performed for an angular range
of $50^\circ \le \theta_{\rm lab} \le 105^\circ$ to cover the main
transfer channels in grazing reactions, whereas the TDHF results are
obtained by Eq.~(\ref{Eq:sigmatot}) without filtering by the scattering
angle, as in Fig.~\ref{FIG:sigma_NZ}.

From the figure, we find that the TDHF results reasonably reproduce
measured cross sections for (0p), ($\pm 1$p), and ($-2$p) channels.
As the number of removed protons increases (($-x$p) with $x \ge 3$),
the peak position of experimental cross sections shifts toward less mass
(neutron) numbers, which is not reproduced by the TDHF results.
The disagreement may partly be originated from the evaporation effect.
We note that, although the peak position is different, the height of the
peaks of the cross sections for proton-stripping channels is in good
agreement with the experimental data, up to ($-4$p) channels.

Similar disagreement was observed in lighter systems
\cite{KS_KY_MNT,KS_KY_FUSION14,Bidyut(2015),KS_KY_Maruhn}.
In Refs.~\cite{KS_KY_FUSION14,Bidyut(2015),KS_KY_Maruhn},
we investigated particle evaporation effects on MNT cross sections
using a statistical model. Although the inclusion of the evaporation
effect improved the agreement between TDHF results and measurements,
there remain disagreements for channels involving a number of transferred
nucleons far from the average values in the TDHF calculation. This failure
indicates a necessity of descriptions beyond the TDHF theory. Beyond
mean-field theories such as the method of Balian and V\'en\'eroni
\cite{BV(1981),Simenel(BV2011)} and the stochastic mean-field
approach \cite{SMF(2008),SMF(2009)1,SMF(2009)2,SMF(2012),
SMF(2013),SMF(2014),SMF(2015)} have recently been developed,
which are expected to remove the discrepancies.

In Fig.~\ref{FIG:NTCS}, we find that the cross sections depend
rather weakly on the initial orientation of $^{238}$U. Difference
is substantial only for ($-5$p), ($-6$p), and ($+2$p) channels.
Differences in these channels are associated with the different
trends of nucleon transfer. The proton-stripping processes are
originated from an impact parameter region, $b\approx 5\mbox{--}6$~fm,
as shown in Fig.~\ref{FIG:Nave}. From the insets shown in
Fig.~\ref{FIG:Nave}~(b,~d), we find that proton-stripping
processes are favored in the $x$-direction case. This trend results
in the difference in ($-5$p) and ($-6$p) channels. In Fig.~\ref{FIG:Nave},
a gradual change of the average number of nucleons was observed
in the $z$-direction case. This change brings a large contribution to
($+1$p) and ($+2$p) channels from a wide impact parameter range.

From a comparison between cross sections by TDHF and those by
{\small GRAZING}, we find that the TDHF results show a better
overall agreement with experimental data. It is remarkable that
the TDHF calculation provides substantial cross sections for the 
proton-pickup channels. The \GRAZING calculation underestimates 
cross sections for those channels, especially for ($+2$p) channel, 
for which the TDHF calculation overestimates. In Ref.~\cite{Corradi(64Ni+238U)},
it was mentioned that lighter fragments with proton numbers up to
$Z \approx 40$ were observed experimentally, especially at forward
angles, although quantitative cross sections were not shown.
The TDHF calculation provides substantial cross sections for lighter 
fragments with $Z\approx 40$ (cf.~Fig.~\ref{FIG:sigma_NZ}~(a-c)) as a 
result of the QF processes at a small-$b$ region, $b \lesssim 4$~fm.

\subsection{Wilczy\'nski plot}

\begin{figure}[t]
\begin{center}
\includegraphics[width=8.6cm]{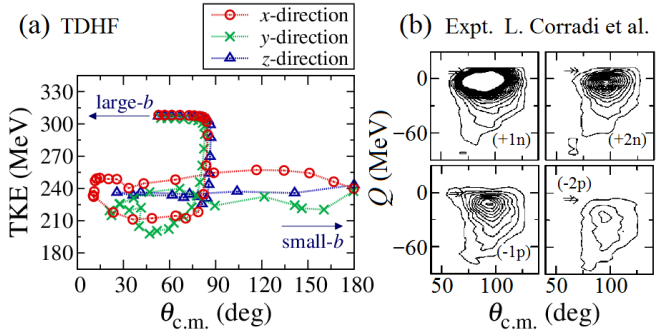}
\end{center}\vspace{-2mm}
\caption{(Color online)
   Wilczy\'nski plots.
   (a): TDHF results for $^{64}$Ni+$^{238}$U reaction at $E_{\rm c.m.}\approx 307.35$~MeV.
   Results for $x\mbox{-},$ $y$-, and $z$-direction cases are shown by red circles,
   green crosses, and blue open triangles connected with dotted lines, respectively.
   (b): Experimental data of  Wilczy\'nski plots for main transfer channels,
   ($+1$n), ($+2$n), ($-1$p), and ($-2$p), in $^{64}$Ni+$^{238}$U
   reaction at $E_{\rm c.m.}\approx 307.35$~MeV. The figures shown in (b) are
   taken from Ref.~\cite{Corradi(64Ni+238U)}.
}
\label{FIG:Wilczynski}
\end{figure}

Combining $\theta_{\rm c.m.}$ in Fig.~\ref{FIG:theta+TKEL_vs_b}~(a)
and TKEL in Fig.~\ref{FIG:theta+TKEL_vs_b}~(b), we obtain the Wilczy\'nski
plot which is shown in Fig.~\ref{FIG:Wilczynski}~(a). In Ref.~\cite{Corradi(64Ni+238U)},
experimental data of Wilczy\'nski plots for various transfer channels were
reported for grazing reaction of $^{64}$Ni+$^{238}$U at $E_{\rm c.m.}
\approx 307.35$~MeV. In (b), we show the experimental data for main transfer
channels, ($+1$n), ($+2$n), ($-1$p), and ($-2$p).

The experimental data show a peak at around $\theta_{\rm c.m.}
\approx 90^\circ$, which shifts toward $Q \approx -60$~MeV (lower TKE),
as the number of transferred protons increases. Our results agree
with the observed trend. At the scattering angle of $\theta_{\rm c.m.}
\approx 80^\circ \mbox{--}85^\circ$ ($b \approx 5\mbox{--}6$~fm),
our TDHF calculation describes proton-stripping processes, as shown
in Sec.~\ref{Sec:Elab390}. In this regime, two nuclei touched gently
at the distance of closest approach, forming a subtle neck which
persists only for a short period. Due to the formation of the subtle neck,
nucleons are exchanged between the projectile and target nuclei
and the TKE decreases rapidly, while the scattering angle is kept
almost constant, $\theta_{\rm c.m.}\approx 80^\circ \mbox{--}85^\circ$. 
In the experimental data shown in (b), we find a tail of the yields which
extends toward forward angles up to $\theta_{\rm c.m.}\approx 75^\circ$,
as the energy loss increases up to $Q\approx -75$~MeV. In the TDHF
results shown in (a), a similar trend is observed at $\theta_{\rm c.m.}
\approx 75^\circ$.

\subsection{Mass-angle distribution}

QF processes are known to show a characteristic correlation between
fragment masses and scattering angle, and thus, MADs in QF processes
have been measured extensively \cite{Toke(QF1985),Bock(1982),duRietz(2011),
duReitz(2013),Wakhle(QF2014)Interplay,Hammerton(2015)}. The MAD
for $^{238}$U+$^{64}$Ni reaction at $E_{\rm c.m.}\approx 302.62$~MeV
was reported by J.~T{\=o}ke et~al.~\cite{Toke(QF1985)}. We compare
the TDHF results with the experimental data. Similar comparisons of MADs
between TDHF calculations and measurements have been reported for
$^{40}$Ca+$^{238}$U \cite{Wakhle(QF2014)Interplay} and
$^{50,54}$Cr+$^{180,186}$W \cite{Hammerton(2015)}.

\begin{figure}[t]
   \begin{center}
   \includegraphics[width=8.6cm]{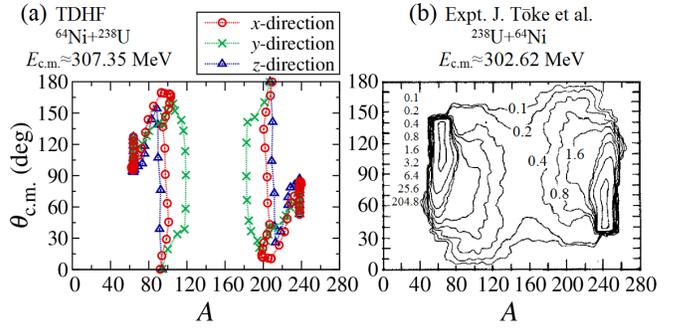}
   \end{center}\vspace{-2mm}
   \caption{(Color online)
   Mass-angle distribution (MAD) plots.
   (a): TDHF results for $^{64}$Ni+$^{238}$U reaction at $E_{\rm c.m.}
   \approx 307.35$~MeV. Results for $x$-, $y$-, and $z$-direction cases are
   shown by red circles, green crosses, and blue open triangles connected
   with dotted lines, respectively. (b): Experimentally measured MAD in
   $^{238}$U+$^{64}$Ni reaction at $E_{\rm c.m.} \approx 302.62$~MeV.
   The figure shown in (b) is taken from Ref.~\cite{Toke(QF1985)}.
   }
   \label{FIG:MAD}
\end{figure}

In Fig.~\ref{FIG:MAD}~(a), we show the MAD plot in the TDHF calculation,
which is constructed from the results shown in Figs.~\ref{FIG:theta+TKEL_vs_b}~(a)
and \ref{FIG:Nave} for $^{64}$Ni+$^{238}$U reaction at $E_{\rm c.m.}
\approx 307.35$~MeV. In (b), the measured MAD for $^{238}$U+$^{64}$Ni
reaction at $E_{\rm c.m.}\approx 302.62$~MeV is shown. Because the inverse
kinematics was employed in the experiment, the angle of $180^\circ
-\theta_{\rm c.m.}$ is used in the plot of (a).

As seen from Figs.~\ref{FIG:theta+TKEL_vs_b}~(a) and \ref{FIG:Nave},
both scattering angle and fragment masses change substantially when
two nuclei start to overlap in the course of collision. These trends
induce correlated behavior in (a) showing an oblique distribution
from $A_{\rm L(H)}=64$ (238) to $A_{\rm L(H)}\approx 100$ (200).
We note that the TDHF calculation provides no contributions to $\theta
_{\rm c.m.}\lesssim$ ($\gtrsim$) $90^\circ$ at $A \approx 64$ (238),
due to the classical nature of the trajectory.

In our TDHF calculations, collisions at a small-$b$ region in which the
mass-drift mode toward the mass symmetry is observed contribute to
certain fragment masses. Reactions of the $x$- and $z$-direction cases 
produce fragments of $A_{\rm L}\approx 90\mbox{--}100$ and $A_{\rm H}
\approx 200\mbox{--}210$, while those of the $y$-direction case produce
fragments around $A_{\rm L}\approx 120$ and $A_{\rm H}\approx 180$. 
Therefore, we expect that the yields in the MAD of the fragments with
symmetric masses are caused by the collisions in the $y$-direction case.

\subsection{Mass-TKE distribution}

\begin{figure}[t]
   \begin{center}
   \includegraphics[width=8.6cm]{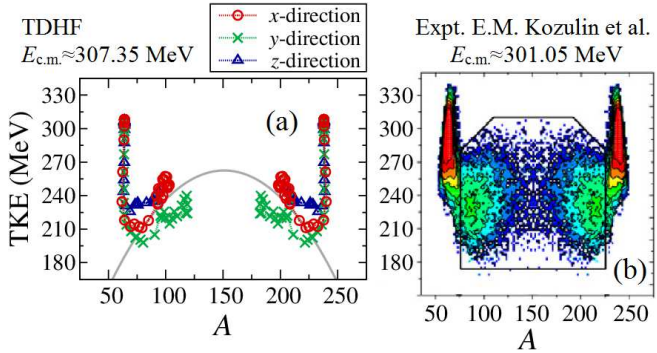}
   \end{center}\vspace{-4mm}
   \caption{(Color online)
   Mass-TKE (total kinetic energy) distribution plots.
   (a): TDHF results for $^{64}$Ni+$^{238}$U reaction at $E_{\rm c.m.}
   \approx 307.35$~MeV. Results for $x$-, $y$-, and $z$-direction cases are
   shown by red circles, green crosses, and blue open triangles connected
   with dotted lines, respectively. The Viola systematics \cite{Viola0,Viola1,Viola2}
   is also shown by a gray solid curve. (b): Experimentally measured mass-TKE
   distribution for $^{64}$Ni+$^{238}$U reaction at $E_{\rm c.m.}=301.05$~MeV.
   The figure shown in (b) is taken from Ref.~\cite{Kozulin(64Ni+238U)}.
   }
   \label{FIG:TKE-A}
\end{figure}

The TKE of outgoing fragments is also a characteristic observable
of QF processes. In Fig.~\ref{FIG:TKE-A}~(a), we show the mass-TKE
distribution in the TDHF calculation, which is constructed from the results
shown in Figs.~\ref{FIG:theta+TKEL_vs_b}~(b) and \ref{FIG:Nave}.
In (b), the measured mass-TKE distribution for $^{64}$Ni+$^{238}$U
reaction at $E_{\rm c.m.}\approx 301.05$~MeV \cite{Kozulin(64Ni+238U)}
is shown. In these plots, two prominent peaks at around $A_{\rm L}
\approx 64$ and $A_{\rm H}\approx 238$ are seen, which correspond to
(quasi)elastic scattering followed by deep-inelastic collisions.
Between these two peaks, QF processes characterized by the
mass-drift mode toward the mass symmetry accompanying
large energy losses are seen.

From the figure, we find a reasonable agreement between the
TDHF results and the experimental data. The TDHF results show
minima at $A_{\rm L}\approx 75$ and $A_{\rm H} \approx 225$ with
TKE~$\approx 200$ in the $y$-direction case. In the measurement,
we find minima at similar values of masses and TKE in the green
areas which correspond to relatively large yields. The mass-TKE
distribution in the TDHF calculation extends toward the mass symmetry,
up to $A_{\rm L}\approx 100\mbox{--}120$ and $A_{\rm H}\approx
180\mbox{--}200$. This trend also reasonably agrees with the 
experimental data in (b).

Both the mass-TKE distributions obtained from the TDHF calculations
and the measurement show a similar trend that the TKE value increases
as the system approaches to the mass symmetry. This behavior can
be understood by the Viola systematics \cite{Viola0,Viola1,Viola2},
which assumes that the TKE is produced by the Coulomb repulsion
at a scission configuration of the composite system. The Viola
systematics is shown by a gray solid curve in (a). Similar comparison
was reported for side collisions of $^{40}$Ca+$^{238}$U at
$b=0$~fm \cite{UO(QF2014)}.

Looking in detail at results shown in (a), our TDHF results indicate
that the $x$-direction case shows smaller dissipation (larger TKE)
than the $y$-direction case, as observed in Fig.~\ref{FIG:theta+TKEL_vs_b}~(b).
This may reflect the shell effect of $^{208}$Pb, which is important
in the $x$-direction case as shown in Fig.~\ref{FIG:Nave}.

\section{SUMMARY AND CONCLUSION}{\label{Sec:summary}}

In this article, we have reported a detailed investigation of
multinucleon transfer (MNT) and quasifission (QF) processes
in $^{64}$Ni+$^{238}$U reaction within the microscopic
framework of the time-dependent Hartree-Fock (TDHF) theory.
For this reaction, abundant experimental data are available for 
both MNT \cite{Corradi(64Ni+238U)} and QF processes 
\cite{Toke(QF1985),Kozulin(64Ni+238U)}. We compared
our  TDHF results with the experimental data.

Because $^{238}$U is substantially deformed in a prolate shape,
we performed TDHF calculations for three initial orientations of
$^{238}$U: the symmetry axis of $^{238}$U is set parallel to
the collision axis, set parallel to the impact parameter vector,
and set perpendicular to the reaction plane. We have called these
three cases as $x$-, $y$-, and $z$-direction cases, respectively.

In all cases, a fast charge equilibration process was observed because
of the large $N/Z$ asymmetry (1.29 for $^{64}$Ni and 1.59 for
$^{238}$U). In peripheral collisions, MNT processes take place.
Several neutrons and protons are transferred toward the direction
expected from the initial $N/Z$ ratios. When two nuclei collide deeply
at small impact parameters, a dinuclear system connected by a thick
neck is formed. When the neck raptures after the charge equilibration
in the dinuclear system, the shape evolution dynamics leading to
the rapture of the neck characterizes the amount of nucleon transfers.
We have regarded the latter process as a QF process, characterized by
a large number of transferred nucleons, a large amount of energy loss, 
and a long contact time. 

For the MNT process, we evaluated transfer cross sections using
the particle-number projection method and compared them with
experimental data \cite{Corradi(64Ni+238U)}. When the number of
transferred nucleons is small, the TDHF calculation reasonably describes
the process. As the number of transferred protons increases, there is
a disagreement between the TDHF calculation and the measurements.
The TDHF calculation gives a peak in the MNT cross section at less 
transferred neutrons than the measurements. Similar success and
failure have been observed in lighter systems \cite{KS_KY_MNT,
Bidyut(2015),KS_KY_FUSION14,KS_KY_Maruhn}.

In QF processes, we have found that there appear substantial orientation
dependence. In $x$- ($y$-) direction case, we have found a larger (smaller)
TKE and a shorter (longer) contact time. In the $x$-direction case
(close to the tip collision), an elongated dinuclear system is formed.
This elongated system tends to split into two fragments in which
a heavier fragment is close to $^{208}$Pb. This indicates a significant
effect of the quantum shells of $^{208}$Pb in the $x$-direction case.
The observed larger TKE and shorter contact time for the $x$-direction
case is thus expected to be originated from the shell effect of $^{208}$Pb
in the QF process. On the other hand, in the $y$-direction case (close to
the side collision), a compact dinuclear system is formed. This system tends
to split into more mass-symmetric fragments, showing less influence of the
quantum shells of $^{208}$Pb. Instead, we have found an indication of
a shell effect of $Z=50$ in the $y$-direction case.

For QF processes, we compared average fragment masses, scattering
angle, and total kinetic energy (TKE) of outgoing fragments with
experimental data \cite{Toke(QF1985),Kozulin(64Ni+238U)}.
From the comparison, we have found that the TDHF calculations reasonably
describe gross behavior seen in the experimental data. We consider
that these agreements are noteworthy since no empirical parameters
are involved in our calculations.

One of the important applications of the microscopic TDHF theory
is to predict optimal conditions to produce objective unstable nuclei,
including superheavy elements. One interesting observation in the
present study is the occurrence of a capture process in the $y$-direction
case at incident energies substantially higher than the Coulomb barrier.
To make a reliable prediction, it is important to take account of the effect
of particle evaporation in competition with fission employing a statistical
model. A study along this direction is in progress \cite{KS_KY_PNP,
KS_KY_FUSION14,KS_KY_Maruhn}.

The present study elucidates to what extent the microscopic TDHF
theory can describe damped collisions of heavy nuclei at low energies,
taking $^{64}$Ni+$^{238}$U system as an example. The significance
of the nuclear structure effects in QF processes is clearly demonstrated
for this system. To increase reliability of descriptions for MNT and QF
processes, extensions of the theory going beyond the mean-field
description are required.

\begin{acknowledgments}
One of the authors (K.S.) appreciates K.~Washiyama and B.J.~Roy
for giving valuable comments on this article. K.S. acknowledges
support of Polish National Science Centre (NCN) Grant, decision
no.~DEC-2013/08/A/ST3/00708. This research used computational
resources of the HPCI system provided by Information Initiative Center,
Hokkaido University, through the HPCI System Research Projects (Project
IDs: hp140010 and hp150081). A part of computation was carried out
at the Yukawa Institute Computer Facility. This work was supported by
the Japan Society for the Promotion of Science (JSPS) Grants-in-Aid for
Scientific Research Grant Numbers 23340113 and 15H03674, and by
the JSPS Grant-in-Aid for JSPS Fellows Grant Number 25-241.
\end{acknowledgments}

\end{document}